\documentclass[12pt,showpacs,preprintnumbers,superscriptaddress,amsmath,amssymb,nofootinbib]{revtex4}
\usepackage{graphicx}
\usepackage{dcolumn}
\usepackage{bm}
\usepackage{amssymb}
\usepackage{amsmath}
\usepackage{epsfig}    
\usepackage{color}
\usepackage{slashed}
\usepackage{hhline}

\def\be{\begin{equation}}
\def\ee{\end{equation}}
\newcommand{\bea}{\begin{eqnarray}}
\newcommand{\eea}{\end{eqnarray}}
\newcommand{\nn}{\nonumber}

\numberwithin{equation}{section}

\begin{document}

\title{Radiative Lepton Model and Dark Matter\\ with Global $U(1)'$
Symmetry} 
\preprint{KIAS-P14003}
\preprint{IPPP/14/07}
\preprint{DCPT/14/14}

\author{Seungwon Baek}
\email{swbaek@kias.re.kr}
\affiliation{School of Physics, KIAS, Seoul 130-722, Korea}
\author{Hiroshi Okada}
\email{hokada@kias.re.kr}
\affiliation{School of Physics, KIAS, Seoul 130-722, Korea}
\author{Takashi Toma}
\email{takashi.toma@durham.ac.uk}
\affiliation{Institute for Particle Physics Phenomenology University of
Durham, Durham DH1 3LE, UK}

\begin{abstract}
We propose a radiative lepton model, in which the charged lepton masses
 are generated at one-loop level, and the neutrino masses are induced at
 two-loop level. On the other hand, tau mass is derived at tree level
 since it is too heavy to generate radiatively. Then we discuss muon
 anomalous magnetic moment together with the constraint of lepton flavor
 violation. A large muon magnetic moment is derived due to the vector
 like charged fermions which are newly added to the standard model. 
 In addition, considering a scalar dark matter in our model, 
 a strong gamma-ray signal is produced by dark matter annihilation via
 internal bremsstrahlung. We can also obtain the effective neutrino
 number by the dark radiation of the Goldstone boson coming from the
 imposed global $U(1)'$ symmetry.  
\end{abstract}
\maketitle
\newpage

\section{Introduction}
Even though 26.8 \% of energy density of our Universe is occupied by  a
non-baryonic dark matter (DM)~\cite{Komatsu:2010fb, Ade:2013zuv}, 
several current  experiments are still under investigation of its nature
from various points of view such as direct and indirect searches. 
As for the direct detection search, for example,
XENON100~\cite{Aprile:2012nq} and LUX~\cite{Akerib:2013tjd} provides the
most severe constraint on spin independent elastic cross section
with nuclei; that is, the cross sections is less than around $10^{-46}$
cm$^2$ at 100 GeV scale of DM mass. 
As for the indirect searches,
AMS-02 has recently shown the positron excess with smooth curve in the
cosmic ray, and reached the energy up to 350 GeV~\cite{ams-02}. 
This result has a good statistics and support the previous experiment
PAMELA~\cite{Adriani:2008zr}. 
On the other hand, 
the recent analysis of gamma-ray observed by Fermi-LAT tells us that
there may be some peak near 130 GeV~\cite{Bringmann:2012vr, Weniger:2012tx}. 
As for the neutrinos, their small masses and mixing pattern call for new
physics beyond the standard model (SM). 
Plank, WMAP9 and ground-based data recently reported a possible deviation in
the effective neutrino number, $\Delta N_{\rm eff}=0.36\pm0.34$ at 68 \%
confidential level~\cite{Ade:2013zuv, Bennett:2012zja,
Das:2013zf, Reichardt:2011yv}. 
Compensating this deviation theoretically might come into one of the
important issues. 
In this sense, radiative seesaw models which support a strong
correlation between DM and neutrinos come into an elegant motivation.  
Many authors have proposed such kind of models in, {\it e.g.},
ref.~\cite{Ma:2006km, Aoki:2013gzs, Dasgupta:2013cwa, Krauss:2002px,
Aoki:2008av, Schmidt:2012yg, 
Bouchand:2012dx, Aoki:2011he, Farzan:2012sa,
Bonnet:2012kz, Kumericki:2012bf, Kumericki:2012bh, Ma:2012if, Gil:2012ya,
Okada:2012np, Hehn:2012kz, Dev:2012sg, Kajiyama:2012xg, Okada:2012sp,
Aoki:2010ib, Kanemura:2011vm, Lindner:2011it, Kanemura:2011mw,
Kanemura:2012rj, Gu:2007ug, Gu:2008zf, Gustafsson, Kajiyama:2013zla, Kajiyama:2013rla,
Hernandez:2013dta, Hernandez:2013hea, McDonald:2013hsa, Okada:2013iba,
Baek:2013fsa, Ma:2014cfa} 
\footnote{Radiative models of the lepton mass are sometimes discussed with
Non-Abelian discrete symmetries due to their selection rules. 
See for example such kind of models:
~\cite{Ahn:2012cg, Ma:2012ez, Kajiyama:2013lja, Kajiyama:2013sza,
Ma:2013mga}.}.

In our paper, we propose a model that neutrino masses as well as charged
lepton (muon and 
electron) masses are generated by radiative correction. 
We obtain a large contribution to muon anomalous magnetic moment from the
charged lepton sector as can be seen later. 
At the same time, one should mind constraints from lepton flavor
violations (LFVs) like $\mu\to e \gamma$ since it is closely
correlated with anomalous magnetic moment. 
Since neutrino masses are generated at two-loop level, they are
therefore naturally suppressed. 
As a result, unlike the TeV scale canonical seesaw mechanism, extremely
small parameters are not required to lead the 
observed neutrino mass scale. 
Moreover the particles run in the loop can be DM candidates. 
Our scalar DM interacts with vector like charged fermions, which are
added to the SM, and the other interaction should be suppressed to
satisfy the direct search constraint. 
Due to the interaction with the vector like charged fermions, a strong
gamma-ray signal is emitted by the DM annihilation via internal
bremsstrahlung preserving consistency with the thermal 
relic density of DM~\cite{Toma:2013bka, Giacchino:2013bta}. In
particular it is possible to adapt with the 
gamma-ray anomaly found in the Fermi data at around 130 GeV. 
The neutrino effective number is also led without conflicting with the
other parts of DM physics. 

This paper is organized as follows.
In Sec.~II, we show our model building for the lepton sector, and
discuss Higgs sector, muon anomalous magnetic moment, and LFV. 
In Sec.~III, DM phenomenology such as relic density, strong gamma-ray
signal and the neutrino effective number is discussed. We summarize and
conclude in Sec.~IV. 

\section{The Model}
\subsection{Model setup}

\begin{table}[thbp]
\centering {\fontsize{10}{12}
\begin{tabular}{|c||c|c|c|c|c|c|c|c|}
\hline Particle & $L_i $ & $ e_{j}^c$  & $ e_{3}^c $  &  $e'_i $
 & $e'^c_i $ & $n'_j $ & $n'^c_j $  & $N^c$ 
  \\\hhline{|=#=|=|=|=|=|=|=|=|}
$(SU(2)_L,U(1)_Y)$ & $(\bm{2},-1/2)$ & $(\bm{1},1)$  & $(\bm{1},1)$ &
		 $(\bm{1},-1)$ & $(\bm{1},1)$ & $(\bm{1},0)$ &
			     $(\bm{1},0)$  & $(\bm{1},0)$  
\\\hline
$U(1)'\times Z_2$ & $(\ell,-)$ & $(0,-)$ & $(-\ell,-)$ & $(\ell,+)$ &
		     $(-\ell,+)$ & $(\ell,+)$ & $(-\ell,+)$  & $(0,-)$
				 \\\hline 
\end{tabular}%
} \caption{The particle contents and the charges for fermions. The $i,j$
 are generation indices: $i=1,2,3$, $j=1,2$.} 
\label{tab:1}
\end{table}

\begin{table}[thbp]
\centering {\fontsize{10}{12}
\begin{tabular}{|c||c|c|c|c|}
\hline Particle   & $\Phi$  & $\eta$  & $\chi$   & $\Sigma$ 
  \\\hhline{|=#=|=|=|=|}
$(SU(2)_L,U(1)_Y)$  & $(\bm{2},1/2)$  & $(\bm{2},1/2)$   & $(\bm{1},0)$
	     & $(\bm{1},0)$ \\\hline 
$U(1)'\times Z_2$  & $(0,+)$ & $(0,-)$ & $(-\ell,-)$  & $(\ell,+)$  \\\hline
\end{tabular}%
} \caption{The particle contents and the charges for bosons. }
\label{tab:2}
\end{table}

We construct a radiative lepton model with global $U(1)'$ symmetry, in
which charged lepton sector is obtained through one-loop level, and
two-loop level for neutrino sector. 
In the model, only tau mass is generated at tree level, but
electron and muon masses are generated at one-loop level. This is
because tau mass is too heavy to generate radiatively. 
The particle contents are shown in Tab.~\ref{tab:1} and
Tab.~\ref{tab:2}. 
The quantum number $\ell(\neq 0)$ in the tables is arbitrary.  
Here $L_i$ and $e^c_i~(i=1,2,3)$ are the SM left-handed and
right-handed lepton fields. 
For right-handed charged leptons $e^c_i~(i=1,2,3)$, different charges of
$U(1)'$ are assigned to the first, second generation and the third
generation in order to distinguish the mass generation mechanism. 
We add three generation of $SU(2)_L$ singlet vector like charged fermions
$e'_i$ and $e'^c_i~(i=1,2,3)$, two generation of vector like neutral
fermions $n'_j$ and $n'^c_j~(j=1,2)$, a singlet 
Majorana fermion $N^c$\footnote{Multi-component vector like fermions are
required to produce the observed charged lepton masses and neutrino
oscillation data. There are  other patterns of 
particle content to derive proper lepton masses.}. 
For new bosons, we introduce $SU(2)_L$ doublet scalar $\eta$ and singlet
scalars 
$\chi$ and $\Sigma$ in addition to the SM Higgs doublet $\Phi$. 
The SM Higgs $\Phi$ should be neutral under $U(1)'$ not to couple quarks to
Goldstone boson through chiral anomaly to be consistent with the axion
particle search\footnote{If $\Phi$ is charged under 
$U(1)'$, its breaking scale should be very large ($\gtrsim 10^{12}$
GeV), which is inconsistent with the observed value $\sim 246$ GeV.}.
We assume that only the SM Higgs doublet $\Phi$ and the SM singlet
$\Sigma$ have vacuum expectation values. 
Otherwise the $\mathbb{Z}_2$ symmetry which guarantees DM stability is
spontaneously broken. 

The renormalizable Lagrangian for Yukawa sector and scalar potential
are given by
\begin{eqnarray}
\mathcal{L}_{Y}
&=&
y_n^{\eta} n^{'c} L  \eta + y_n^{\chi} N^c n' \chi
+ \frac{M_N}{2} N^c N^c + M_{n'} n'^c n' +\rm{h.c.}\nn\\ 
&&+
y_\tau^{\Phi } \Phi^\dag e_{3}^c L +
y_\ell^{\eta} \eta^\dag  e'^c L + y_\ell^{\chi} e_{1,2}^c e' \chi +M_{e'}
e' e'^c +\rm{h.c.} \\ 
\mathcal{V}
&=& 
 m_1^{2} \Phi^\dagger \Phi + m_2^{2} \eta^\dagger \eta  + m_3^{2}
 \Sigma^\dagger \Sigma  + m_4^{2} \chi^\dagger \chi
  +\lambda_1 (\Phi^\dagger \Phi)^{2} + \lambda_2 
(\eta^\dagger \eta)^{2} + \lambda_3 (\Phi^\dagger \Phi)(\eta^\dagger \eta) 
\nn\\
&&+
 \lambda_4 (\Phi^\dagger \eta)(\eta^\dagger \Phi)
+
\lambda_5 [(\Phi^\dagger \eta)^{2} + \mathrm{h.c.}]+
\lambda'_5 [(\Sigma^\dagger \chi)^{2} + \mathrm{h.c.}]+
\lambda''_5 [(\Sigma \chi)^{2} + \mathrm{h.c.}]
\nn\\&&+
\lambda_6 (\Sigma^\dagger \Sigma)^{2} +\lambda'_6 (\chi^\dagger \chi)^{2} +
\lambda''_6 \left(\Sigma^\dagger\Sigma\right)\left(\chi^\dag\chi\right) 
+
\lambda_7  (\Sigma^\dagger \Sigma)(\Phi^\dagger \Phi ) + \lambda'_7
(\chi^\dagger \chi)(\Phi^\dagger \Phi) \nn\\
&&+ \lambda_8  (\Sigma^\dagger \Sigma) (\eta^\dagger \eta)+ \lambda'_8
 (\chi^\dagger \chi) (\eta^\dagger \eta) 
+\left[a(\eta^\dag \Phi)(\Sigma\chi)+{\rm
  h.c.}\right]+\left[a'(\Phi^\dag \eta)(\Sigma\chi)+{\rm h.c.}\right],
\nonumber\\
\label{HP}
\end{eqnarray}
where $\lambda_5$, $\lambda_5'$, $\lambda_5''$, and one of $a$ and $a'$
can be chosen to be real 
without any loss of generality by absorbing the phases to scalar bosons. 
The $\Phi^\dagger e_i^{\prime c} L$ term which might generate mixing
between $e_i^{\prime c}$ and 
$e_3^c$ is not allowed by the $\mathbb{Z}_2$ symmetry. The Yukawa interaction
$\Phi^\dagger e_{1,2}^c L$ 
which gives the tree level masses of electron and muon is forbidden by
$U(1)'$ symmetry. 
The term $N^c L \eta$ which induces one-loop neutrino
masses~\cite{Ma:2006km} is also excluded by $U(1)'$ symmetry. 
The couplings $\lambda_1$, $\lambda_2$, $\lambda_6$ and 
$\lambda'_6$ have to be positive
to stabilize the Higgs potential.  
Inserting the tadpole conditions; $m^2_1=-\lambda_1v^2-\lambda_7v'^2/2$ and
 $m^2_3=-\lambda_6v'^2 - \lambda_7v^2/2$,
the resulting mass matrix of the neutral component of $\Phi$ and 
$\Sigma$ defined as 
\begin{equation}
\Phi^0=\frac{v+\phi^0(x)}{\sqrt{2}},\qquad
\Sigma=\frac{v'+\sigma(x)}{\sqrt{2}}e^{iG(x)/v'},\quad
\end{equation}
 is given by
\begin{equation}
m^{2} (\phi^{0},\sigma) = \left(%
\begin{array}{cc}
  2\lambda_1v^2 & \lambda_7vv' \\
  \lambda_7vv' & 2\lambda_6v'^2 \\
\end{array}%
\right) \!=\! \left(\begin{array}{cc} \cos\alpha & \sin\alpha \\ -\sin\alpha & \cos\alpha \end{array}\right)
\left(\begin{array}{cc} m^2_{h} & 0 \\ 0 & m^2_{H}  \end{array}\right)
\left(\begin{array}{cc} \cos\alpha & -\sin\alpha \\ \sin\alpha &
      \cos\alpha \end{array}\right), 
\end{equation}
where $h$ implies SM-like Higgs with the mass of 125 GeV and $H$ is an
additional CP-even Higgs mass eigenstate. The mixing angle $\alpha$ is
given by 
\be
\tan 2\alpha=\frac{\lambda_7 v v'}{\lambda_6 v'^2-\lambda_1 v^2}.
\ee
The Higgs bosons $\phi^0$ and $\sigma$ are rewritten in terms of the
mass eigenstates $h$ and $H$ as 
\begin{eqnarray}
\phi^0 &=& h\cos\alpha + H\sin\alpha, \nn\\
\sigma &=&- h\sin\alpha + H\cos\alpha.
\label{eq:mass_weak}
\end{eqnarray}
A Goldstone boson $G$ appears due to the spontaneous symmetry breaking of
the global $U(1)'$ symmetry. 
This massless particle would be dark radiation contributing to
the effective neutrino number we will discuss later~\cite{Weinberg:2013kea}.

The resulting mass matrix of the neutral component of $\eta$ and 
$\chi$ defined as 
\begin{equation}
\eta^0=\frac{\eta_R+i\eta_I}{\sqrt{2}},\qquad
\chi=\frac{\chi_R+i\chi_I}{\sqrt{2}},\quad
\end{equation}
 is given by
\begin{equation}
m^{2} (\eta_R,\chi_R) = \left(%
\begin{array}{cc}
m^2_{\eta_{R}} & m^2_{\eta_R\chi_{R}} \\
m^2_{\eta_R\chi_{R}} & m^2_{\chi_{R}} \\
\end{array}%
\right) \!=\! \left(
\begin{array}{cc} 
 \cos\beta_R & \sin\beta_R \\
-\sin\beta_R & \cos\beta_R 
\end{array}
\right) 
\left(\begin{array}{cc} m^2_{h'_R} & 0 \\ 0 & m^2_{H'_R}  \end{array}\right)
\left(
\begin{array}{cc} 
\cos\beta_R & -\sin\beta_R \\ 
\sin\beta_R &  \cos\beta_R 
\end{array}
\right),  
\end{equation}
for CP even mass eigenstates where $h_R'$ and $H_R'$ are mass
eigenstates of inert Higgses.
The imaginary part of these inert Higgses (CP odd states) is defined by replacing the index $R$ into $I$, hereafter.
The mixing angle $\beta_R$ is given by 
\be
\tan 2\beta_R=\frac{2m^2_{\eta_R\chi_{R}} }
{ m^2_{\chi_{R}} -m^2_{\eta_{R}} }.
\ee
The $\eta_R$ and $\chi_R$ are rewritten in terms of the mass eigenstates
$h'_R$ and $H'_R$ as 
\begin{eqnarray}
\eta_R &=& h'_R\cos\beta_R + H'_R\sin\beta_R, \nn\\
\chi_R &=&- h'_R\sin\beta_R + H'_R\cos\beta_R.
\label{eq:mass_weak}
\end{eqnarray}
Each mass component is defined as
\begin{eqnarray}
m_\eta^2\equiv m^{2} (\eta^{\pm}) &=& m_2^{2} + \frac12 \lambda_3 v^{2}
 + \frac12 \lambda_8 v'^{2}, \\ 
m^2_{\eta_{R}}\equiv m^{2} (\eta_R) &=& m_2^{2} + \frac12
 \lambda_8 v'^{2} 
 + \frac12 (\lambda_3 + \lambda_4 + 2\lambda_5) v^{2}, \\ 
m^2_{\eta_{I}}\equiv m^{2} (\eta_I) &=& m_2^{2} + \frac12
 \lambda_8 v'^{2} 
 + \frac12 (\lambda_3 + \lambda_4 - 2\lambda_5) v^{2},\\
m^2_{\chi_{R}} \equiv m^2(\chi_R)&=& m_3^{2} 
+ \frac12\left(\frac12 \lambda''_6 v'^2 + \frac12\lambda'_{7} v^{2} +
	  \lambda'_5 v'^2+  \lambda''_5 v'^2\right), \\  
m^2_{\chi_{I}} \equiv m^2(\chi_I)&=& m_3^{2} 
+ \frac12\left(\frac12 \lambda''_6 v'^2 + \frac12\lambda'_{7} v^{2} -
	  \lambda'_5 v'^2 -  \lambda''_5 v'^2\right),\\ 
m^2_{\eta_R\chi_{R}} &=& \frac14  vv' (a+a'),\quad  
m^2_{\eta_I\chi_{I}} =  \frac14  vv' (a-a').
\label{eq:scalar-mixing}
\end{eqnarray}
We note that we need mass splitting between $\eta_R (\chi_R)$ and
$\eta_I (\chi_I)$ which is required to generate the non-zero lepton masses. 
The tadpole conditions for $\eta$ and $\chi$, which are given by
$\left.\partial \mathcal{V}/\partial \eta\right|_{\mathrm{VEV}}=0$, 
$\left.\partial \mathcal{V}/\partial \chi\right|_{\mathrm{VEV}}=0$, 
$0<\left.\partial^2 \mathcal{V}/\partial \eta^2\right|_{\mathrm{VEV}}$ and 
$0<\left.\partial^2 \mathcal{V}/\partial \chi^2\right|_{\mathrm{VEV}}$ 
tell us that  
\be
0<m^2_2 + \frac{v^2}2( \lambda_3+\lambda_4+2\lambda_5)+
\frac{v'^2}2\lambda_8,\quad 
0<m^2_4 + \frac{v^2}2 \lambda'_{7} + \frac{v'^2}2
(\lambda'_{5}+\lambda''_{5}+\lambda''_{6}), 
\ee
to satisfy the condition $\langle\eta\rangle=0$ and
$\langle\chi\rangle=0$ at tree level, respectively. 
In order to avoid that $\langle \Phi \rangle=\langle \Sigma \rangle=0$ be 
a local minimum, we require the following condition: 
\be
\lambda_7-\frac{2}{3}\sqrt{\lambda_1\lambda_6}<0.
\ee
To achieve the global minimum  at $\langle \eta \rangle=\langle \chi
\rangle=0$, we find the following condition 
\be
0<\lambda'_{8}-\frac{2}{3}\sqrt{\lambda_2\lambda'_6}.
\ee
Finally, if the following conditions  
\bea
&&0<\lambda_{3}+\frac{2}{3}\sqrt{\lambda_1\lambda_2},
\quad 0<\lambda_{7}+\frac{2}{3}\sqrt{\lambda_1\lambda_6},\quad
0<\lambda'_{7}+\frac{2}{3}\sqrt{\lambda_1\lambda'_6},
\nn\\&&
 0<\lambda_{8}+\frac{2}{3}\sqrt{\lambda_2\lambda_6},\quad
0<\lambda'_{8}+\frac{2}{3}\sqrt{\lambda_2\lambda'_6},\quad 
0<\lambda''_{6}+\frac{2}{3}\sqrt{\lambda_6\lambda'_6}, 
\eea
are satisfied, the Higgs potential Eq.~(\ref{HP}) is bounded from below.

\subsection{Charged lepton and neutrino mass matrix}
\begin{figure}[cbt]
\begin{center}
 \includegraphics[scale=1]{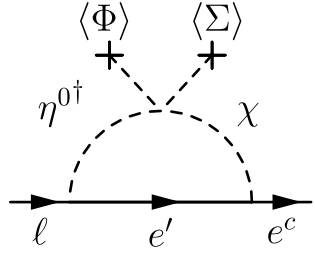}
   \caption{Radiative generation of charged lepton masses.  
   }
   \label{cgd-lepton-diag}
\end{center}
\end{figure}
\begin{figure}[cbt]
\begin{center}
\includegraphics[scale=1]{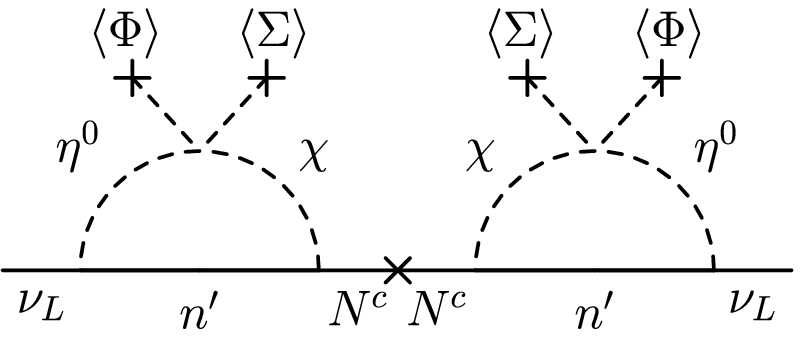}
\qquad
 \includegraphics[scale=1]{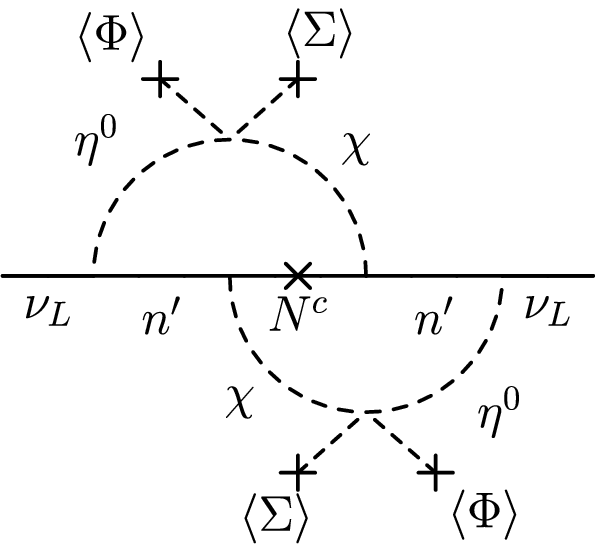}
   \caption{Radiative generation of neutrino masses.  
   }
   \label{neutrino-diag}
\end{center}
\end{figure}

The tau mass is given at tree level, after the spontaneous symmetry
breaking as $m_\tau =y_\tau^\Phi v/\sqrt2$. 
On the other hand, the electron and muon masses are generated at
one-loop, as can be 
seen in Fig.~\ref{cgd-lepton-diag} as follows: 
\be
(m_{\ell})_{\alpha\beta}=\sum_{i}\frac{(y_\ell^\eta)_{\alpha i}
(y_\ell^\chi)_{i\beta}
M_{e'i}\sin2\beta_R}{4(4\pi)^2}
\left[F\left(\frac{m_{h_R'}^2}{M_{e'i}^2}\right)
-F\left(\frac{m_{H_R'}^2}{M_{e'i}^2}\right)\right] 
+ (R\to I), 
\label{eq:lepton-mass}
\ee
where $F(x)=x\log{x}/(1-x)$. 
The total mass matrix is diagonalized by bi-unitary matrix. 
From the mass formula, for example, the Yukawa coupling
$(y_{\ell}^{\eta}y_{\ell}^{\chi})\sim1$ is required 
for muon mass and $(y_{\ell}^{\eta}y_{\ell}^{\chi})\sim0.01$ for
electron mass when $M_{e'}\sim500~\mathrm{GeV}$,
$\sin2\beta_{R(I)}\sim0.1$ and $\mathcal{O}(1)$ of the loop function. 
The Yukawa coupling $y_{\ell}^\chi$ should be $\mathcal{O}(1)$ to obtain the
observed DM relic density as we will see in Sec.~\ref{sec:DM}. 

The Dirac neutrino mass matrix at one-loop level as depicted in
the left hand side of Fig.~\ref{neutrino-diag} is given by 
\begin{equation}
(m_{D})_{i\beta}=
\sum_{i}\frac{(y_n^\chi)_{i}(y_n^\eta)_{i\beta
}M_{n'i}\sin{2\beta_R}}{4(4\pi)^2} 
\left[F\left(\frac{m_{h_R'}^2}{M_{n'i}^2}\right)
-F\left(\frac{m_{H_R'}^2}{M_{n'i}^2}\right)\right]
- (R\to I),
\end{equation}
With the Dirac neutrino mass matrix, the active neutrino mass matrix is
obtained by canonical seesaw mechanism as 
\begin{equation}
(m_{\nu1})_{\alpha\beta}=-\frac{1}{M_N}\left(m_D^Tm_D\right)_{\alpha\beta}.
\end{equation}
In addition, there is another contribution to the neutrino masses coming
from the right hand side of Fig.~\ref{neutrino-diag}. The mass matrix is
expressed as~\cite{Kanemura:2011mw}
\begin{equation}
(m_{\nu2})_{\alpha\beta}=
\sum_{i}\sum_{k}\frac{(y_n^\eta)_{i\alpha
}(y_n^\chi)_i(y_n^\chi)_k(y_n^\eta)_{k\beta
}M_{n'i}M_N}{16(4\pi)^4M_{n'k}}
F_{ik}^{\mathrm{loop}},
\end{equation}
where the loop function $F_{ik}^{\mathrm{loop}}$ is given by
\begin{eqnarray}
F_{ik}^{\mathrm{loop}}&=&
\int d^3x\frac{\delta(x+y+z-1)}{y(y-1)}\nonumber\\
&&\times
\left[
\left\{\sin^22\beta_R\left(
G\left(\frac{M_{ih_R'}^2}{M_{n'k}^2},\frac{m_{h_R'}^2}{M_{n'k}^2}\right)
-G\left(\frac{M_{ih_R'}^2}{M_{n'k}^2},\frac{m_{H_R'}^2}{M_{n'k}^2}\right)
\right)
+\left(h_R'\leftrightarrow H_R'\right)\right\}\right.\nonumber\\
&&\left.\qquad
-\left\{\sin2\beta_R\sin2\beta_I\left(
G\left(\frac{M_{ih_R'}^2}{M_{n'k}^2},\frac{m_{h_I'}^2}{M_{n'k}^2}\right)
-G\left(\frac{M_{ih_R'}^2}{M_{n'k}^2},\frac{m_{H_I'}^2}{M_{n'k}^2}\right)
\right)
-\left(h_R'\leftrightarrow H_R'\right)\right\}\right.\nonumber\\
&&\Biggl.\qquad+(R\leftrightarrow I)
\Biggr],
\end{eqnarray}
with
\begin{equation}
G\left(x,y\right)=\frac{-x(1-y)\log{x}+y(1-x)\log{y}}{(1-x)(1-y)(x-y)},
\end{equation}
and 
\begin{equation}
M_{ia}^2\equiv
\frac{xm_{n'i}^2+yM_N^2+zm_{a}^2}{y(y-1)}
\end{equation}
where $a=h_R',H_R',h_I',H_I'$.
Whole neutrino mass matrix is sum of the two contributions as
$m_{\nu}=m_{\nu1}+m_{\nu2}$. 
From the neutrino mass formula, $(y_{n}^{\chi}y_{n}^{\eta})\sim0.01$ is
needed to obtain the proper neutrino mass scale by assuming
$M_{n'}\sim500~\mathrm{GeV}$, $M_N\sim1~\mathrm{TeV}$,
$\mathcal{O}(0.1)$ of the loop functions.

\subsection{The muon anomalous magnetic moment and Lepton Flavor Violation}
The muon anomalous magnetic moment, $(g-2)_\mu$, has been 
measured at Brookhaven National Laboratory. 
The current average of the experimental results~\cite{bennett} is given by
\begin{align}
a^{\rm exp}_{\mu}=11 659 208.0(6.3)\times 10^{-10},\notag
\end{align}
which has a discrepancy from the SM prediction with
$3.2\sigma$~\cite{discrepancy1} to $4.1\sigma$~\cite{discrepancy2} as 
\begin{align}
\Delta a_{\mu}=a^{\rm exp}_{\mu}-a^{\rm SM}_{\mu}=(29.0 \pm
9.0\ {\rm to}\ 33.5 \pm
8.2)\times 10^{-10}. \notag
\end{align}
In our model, there are several contributions to the (transition)
magnetic moment $\mu_{\alpha\beta}$ which is 
coefficient of the operator
$\mu_{\alpha\beta}\overline{\ell_\alpha}
\sigma^{\mu\nu}\ell_{\beta}F_{\mu\nu}$. 
The muon anomalous magnetic moment is identified as $\Delta
a_{\mu}=\mu_{\mu\mu}$. 
The largest contribution comes from
photon attaching to vector like charged fermions since it is proportional
to $m_{\alpha}/M_{e'}$ where $m_\alpha$ is charged lepton mass. On the
other hand, the other contributions are 
proportional to $m_{\alpha}^2/M_{e'}^2$. 
The contributions coming from the loop of $\eta^+$ and $n'$ in neutrino
sector are also proportional to $m_\alpha^2/M_{n'}^2$. 
Thus these are neglected in our
calculation, and the (transition) magnetic moment is calculated as 
\bea
\mu_{\alpha\beta}&\simeq& \sum_{i=1}^2\frac{ \sin2\beta_R}{2(4\pi)^2 }
\frac{m_\alpha}{M_{e'i}}
\Bigl((y_\ell^{\eta})^{\alpha i} (y_\ell^{\chi})^{i\beta
}+(y_\ell^{\eta*})^{\beta i} (y_\ell^{\chi*})^{i\alpha}\Bigr) 
\left[-H\left(\frac{m_{h_R'}^2}{M_{e'i}^2}\right)
+H\left(\frac{m_{H_R'}^2}{M_{e'i}^2}\right)\right]
+(\rm{R}\to{\rm I}) \
 \nn\\&& 
 {\rm with}
 ~H(x)=\frac{1-4x +3 x^2-2 x^2 \ln x}{2(1-x)^3}. 
\eea
More preciously, the unitary matrices which diagonalize the charged
lepton mass matrix should be multiplied from left and right. It
is understood by replacing Yukawa couplings $y_\ell^\eta$, $y_\ell^\chi$
to $y_\ell^{\eta'}$, $y_\ell^{\chi'}$. 
This expression of the (transition) magnetic moment is closely related with
radiative induced charged lepton masses Eq.~(\ref{eq:lepton-mass}). 
To reproduce the muon mass, for example, 
$\sin2\theta_{R(I)}$ and $M_{e'i}$ are taken to be
$\mathcal{O}$($10^{-2}$) and $\mathcal{O}$(1) TeV, respectively. 
Thus we obtain $\Delta a_{\mu}={\cal O}$(10$^{-9}$), when
$(y_\ell^{\eta})
(y_\ell^{\chi})[H(m_{h_R'}^2/M_{e'i}^2)-H(m_{H_R'}^2/M_{e'i}^2)]$
is roughly 0.1. 

It is the common fact that
muon $g-2$ and Lepton Flavor Violation tend to conflict each other. 
In LFV processes, $\mu\to e\gamma$ especially gives
the most stringent bound. 
The upper limit of the branching ratio is given by
$\mathrm{Br}\left(\mu\to e\gamma\right)\leq5.7\times 10^{-13}$ at 95\%
confidence level from the MEG experiment~\cite{meg}. 
In our model, the diagonal Yukawa matrix 
$y_{\ell}^\eta$ and $y_{\ell}^\chi$ is required not to conflict with
Lepton Flavor Violating processes such as $\mu\to e\gamma$. 
Nevertheless, the contribution to $\mu\to e\gamma$ still comes from the
neutrino sector, and it is calculated as
\begin{equation}
\mathrm{Br}\left(\mu\to e\gamma\right)=
\frac{3\alpha_{\mathrm{em}}}{64\pi G_F^2m_\eta^4}
\left|\sum_{i}\left(y_{n}^\eta\right)_{i\mu}
\left(y_n^{\eta}\right)_{ie}^*F_2\left(\frac{M_{n'i}^2}{m_\eta^2}\right)
\right|^2, 
\label{eq:lfv}
\end{equation}
where $\alpha_{\mathrm{em}}=$1/137 is the fine structure constant, $G_F$
is the Fermi constant and $F_2(x)$ is the loop function defined in
ref.~\cite{Ma:2001mr}. 
From the Eq.~(\ref{eq:lfv}), we obtain a rough estimation for the Yukawa
coupling $y_{n}^\eta\lesssim0.05$ by setting
$m_\eta=M_{n'}\sim500~\mathrm{GeV}$. This estimation does not contradict with
the discussion of neutrino masses.


\section{Dark Matter}
\label{sec:DM}
We have two DM candidates: vector like fermion $n'$, the lightest
eigenstate of $\eta^0$ and $\chi$ (one of $h_R'$, $H_R'$, $h_I'$, $H_I'$). 
One may think the scalar DM candidate decays into the SM particles 
since the SM leptons also have odd charge under the imposed $\mathbb{Z}_2$
symmetry in our model. However, the decay of the DM candidate is
forbidden by Lorentz invariance. Namely, this means 
that the scalar DM candidate can decay into only even number of fermion,
however such a decay process is not allowed in the model.

We identify $h_R'$ is DM here since it has interesting DM
phenomenology. 
The mixing angle $\sin\beta_R$ is needed to be small enough since 
tiny neutrino masses are proportional to the mixing angle. 
Note that in the limit of $\sin\beta_R\to0$, there is no contribution
from $h_R'$ and $H_R'$ to the charged lepton and neutrino 
masses as one can see from the previous section. 
However we still have the contribution of $h_I'$ and $H_I'$. The
neutrino masses are generated from $h_I'$ and $H_I'$. 
The parameter relation $a\approx -a'$ is
required to construct such a situation as one can see
Eq.~(\ref{eq:scalar-mixing}). 
In this case, the DM candidate $h_R'$
corresponds to just $\chi_R$. Thus we regard $\chi_R$ as DM hereafter. 
The couplings $\lambda_5'$, $\lambda_5''$, $\lambda_6''$ and $\lambda_7'$
in the scalar potential also should be suppressed not to
have large elastic cross section with nuclei. 
Otherwise elastic scattering occurs via Higgs exchange and it is
excluded by direct detection experiments of DM 
such as XENON~\cite{Aprile:2012nq} or LUX~\cite{Akerib:2013tjd}. 
The spin independent elastic cross section with proton in the limit
of $\sin\beta_R\to0$ is given by
\begin{equation}
\sigma_p=\frac{C\mu_{\chi}^2m_p^2}{\pi m_{\chi_R}^2v^2}
\left(\frac{\mu_{\chi\chi h}\cos\alpha}{m_h^2}
+\frac{\mu_{\chi\chi H}\sin\alpha}{m_H^2}\right)^2,
\label{eq:elastic}
\end{equation}
where $\mu_\chi$ is reduced mass defined as
$\mu_\chi=(m_{\chi_R}+m_p^{-1})^{-1}$, $m_p=938~\mathrm{MeV}$ is the proton
mass and $C\approx0.079$. 
The couplings $\mu_{\chi\chi h}$ and $\mu_{\chi\chi H}$ are 
given by 
\begin{eqnarray}
\mu_{\chi\chi h}&=&
-\left(\lambda_5'+\lambda_5''+\frac{\lambda_6''}{2}\right)v'\sin\alpha
+\frac{\lambda_7'}{2}v\cos\alpha, \\
\mu_{\chi\chi
H}&=&\left(\lambda_5'+\lambda_5''+\frac{\lambda_6''}{2}\right)v'\cos\alpha
+\frac{\lambda_7'}{2}v\sin\alpha. 
\end{eqnarray}
The elastic cross section is strongly constrained by LUX as $\sigma_p\lesssim
7.6\times10^{-46}~\mathrm{cm^2}$ 
at $m_{\chi_R}\approx33~\mathrm{GeV}$. 
Thus the couplings $\lambda_5'$, $\lambda_5''$, $\lambda_6''$ and
$\lambda_7'$ are required to be $\mathcal{O}(0.001)$ in order to satisfy
the constraint when $v'\sim1~\mathrm{TeV}$ and $\sin\alpha\sim1$. 

Due to the strong constraint from direct detection of DM, 
the annihilation cross section for the process $\chi_R\chi_R\to
f\overline{f}$ via Higgs s-channel is extremely
suppressed. The cross section is calculated as 
\begin{equation}
\sigma{v}_{\mathrm{rel}}=\frac{y_f^2}{2\pi}\left(1-\frac{4m_f^2}{s}\right)^{3/2}
\left|\frac{\mu_{\chi\chi h}\cos\alpha}{s-m_h^2+im_h\Gamma_h}
+\frac{\mu_{\chi\chi H}\sin\alpha}{s-m_H^2+im_H\Gamma_H}\right|^2,
\end{equation}
where $s\approx 4m_{\chi_R}^2(1+v_{\mathrm{rel}}^2/4)$, $\Gamma_h$ and
$\Gamma_H$ are the decay width of $h$ and $H$. 
With the above constraint from direct detection, the typical value of
the annihilation cross section is roughly
$\sigma{v}_{\mathrm{rel}}\sim10^{-32}~\mathrm{cm^3/s}$ which is too 
small to obtain the observed DM relic density $\Omega
h^2\approx0.12$~\cite{Ade:2013zuv}. 

However there is the Yukawa interaction $y_{\ell}^{\chi}e^ce'\chi$. 
The DM annihilation $\chi_R\chi_R\to \ell\overline{\ell}$ is possible via
the Yukawa interaction. 
When one expands the cross section by the DM relative velocity
$v_{\mathrm{rel}}$, the s-wave and p-wave of the process are helicity
suppressed. 
Thus this process becomes d-wave dominant in the chiral limit of the
final state particles as have studied in ref.~\cite{Toma:2013bka, 
Giacchino:2013bta}. The annihilation cross 
section is written as
\begin{equation}
\sigma{v}_{\mathrm{rel}}=\left|\sum_{i}
\frac{(y_{\ell}^{\chi\dag}y_{\ell}^{\chi})_{ii}} 
{(1+\mu_i)^2}\right|^2\frac{v_{\mathrm{rel}}^4}{60\pi m_{\chi_R}^2}, 
\end{equation}
where $\mu_i=m_{e'i}^2/m_{\chi_R}^2>1$. 
The Yukawa couplings should be $\mathcal{O}(1)$ to
achieve the correct relic density of the DM. 
As a result of the d-wave suppression of the 2-body cross section,
internal bremsstrahlung process $\chi_R\chi_R\to
\ell\overline{\ell}\gamma$ which 
generates sharp gamma ray spectrum around $E_\gamma\sim m_\chi$ becomes 
strong as can be compared with the experiments such as
Fermi-LAT~\cite{Fermi-LAT:2013uma} 
or future project CTA~\cite{Consortium:2010bc} without conflicting
with the thermal relic 
density of DM. The predicted spectrum is stronger than that in case of
p-wave dominant Majorana DM~\cite{Bringmann:2012vr}. 
When $\mu_i$ is far from $1$, the gamma ray spectrum becomes broader. Thus 
roughly $\mu_i\lesssim2$ is needed to produce a sharp gamma ray spectrum.

Finally, we mention about the discrepancy of the effective 
number of neutrino species $\Delta N_{\rm eff}$. 
This has been reported by several
experiments such as Planck~\cite{Ade:2013zuv}, WMAP9
polarization~\cite{Bennett:2012zja}, and ground-based
data~\cite{Das:2013zf, Reichardt:2011yv}, 
which tell us $\Delta N_{\rm eff}=0.36 \pm 0.34$ at the 68~\% confidence
level. 
Such a deviation $\Delta N_{\rm eff}\approx0.39$ is
achieved, if we take the extra neutral boson $H$ to be light as well as
$500$ MeV and small mixing angle $\sin\alpha\ll1$~\cite{Weinberg:2013kea,
Garcia-Cely:2013nin, Okada:2013iba}. 
Such a light mass is needed to determine the appropriate decoupling era
of the extra neutral boson in the early Universe. 
The mixing angle also should be small enough to suppress the invisible decay
of the SM Higgs $h\to HH$. When such a light extra Higgs $H$ is
taken into account, smaller scalar couplings $\lambda_5'$, $\lambda_5''$,
$\lambda_6'$ are required to be consistent with the constraint on
elastic cross section with proton Eq~(\ref{eq:elastic}). However 
it does not matter with the estimation of the thermal relic density and
the strong gamma-ray signal discussed above because these are induced
via the Yukawa coupling $y_\ell^\chi$. 
Hence we can derive the neutrino effective number $\Delta N_{\rm
eff}$ without any contradiction with the other DM phenomenology.

\section{Conclusions}
We have constructed a model where the neutrino and charged lepton masses
are generated radiatively. The electron and muon masses are 
obtained from one-loop diagram while the neutrino masses arise through
two-loop diagrams. The tau mass is rather heavy to generate radiatively,
and is given by the tree level 
Yukawa interaction. Thus their measured mass hierarchies are naturally
explained. Then we have obtained the large muon anomalous  
magnetic moment ($(g-2)_\mu$) as same as the observed value from the
charged lepton sector. 
Such a large magnetic moment tends to conflict with LFV processes. 
To avoid this, an appropriate parameter condition have been considered
to be consistent with LFV.

The same symmetries that explain charged lepton and neutrino masses also
allow some DM candidates.
We have shown that our scalar DM can emit a strong gamma-ray by internal
bremsstrahlung process which is possible to compare with the experiment
such as Fermi-LAT. In addition, the thermal relic density of DM can be
consistently derived unlike internal bremsstrahlung of Majorana DM. 
Simultaneously, when $H$ is light ($m_H \sim 500$ MeV) and the mixing
angle $\sin\alpha$ is small enough, the Goldstone boson can play the
role of dark radiation and 
we can also induce a sizable discrepancy in the effective
neutrino number $\Delta N_{\rm eff}\approx 0.39$.

\section*{Acknowledgments}
This work is partly supported by NRF Research Grant  2012R1A2A1A01006053 (SB).
T.T. acknowledges support from the European ITN project (FP7-PEOPLE-2011-ITN,
PITN-GA-2011-289442-INVISIBLES). 

\end{document}